
\magnification=1200
\footline={\ifnum\pageno=1\hfill\else\hfill\rm\folio\hfill\fi}
\font\title=cmr12 at 12pt
\baselineskip=18pt
\def\dsl{\raise.15ex\hbox{/}\kern-.57em\partial}
\def\Dsl{\,\raise.15ex\hbox{/}\mkern-13.5mu D} 
\def\Asl{\,\raise.15ex\hbox{/}\mkern-13.5mu A} 
\def\Bsl{\,\raise.15ex\hbox{/}\mkern-13.5mu B} 
\def\mbox#1#2{\vcenter{\hrule \hbox{\vrule height#2in
                \kern#1in \vrule} \hrule}}
\def\Box {\mbox {.08} {.08}\,}
\vskip 1.0cm
\centerline{{\title ELECTROMAGNETIC INTERACTION }}
\centerline{{\title OF ANYONS IN NON-RELATIVISTIC }}
\centerline{{\title  QUANTUM FIELD THEORY}\footnote{$^\dagger$}{Work
partially supported by CICYT (Proyecto AEN 90-0030).}}
\vskip 2.0cm
\centerline{\bf J. L. Cort\'es$^1$, J. Gamboa$^2$ and
L. Vel\'azquez$^1$}
\centerline{\it $^1$Departamento de F\'\i sica Te\'orica, Universidad de
Zaragoza}
\centerline{\it 50009 Zaragoza, Spain.}
\centerline{\it $^2$ Division de Physique Th\'eorique, Institut de Physique
Nucleaire\footnote{$^\ddagger$}{Unit\'e de Recherche des Universit\'es Paris
11 et Paris 6 associ\'ee au CNRS.}}
\centerline{\it F-91406, Orsay Cedex, France.}
\vskip 2.0cm

{\bf Abstract}.
The non-relativistic quantum field theoretic lagrangian
which describes an anyon system in the presence of an
electromagnetic field is identified. A non-minimal magnetic
coupling to the Chern-Simons statistical field as well as
to the electromagnetic field together with a direct coupling
between both fields are the non trivial ingredients of the
lagrangian obtained from the non relativistic limit of the
fermionic relativistic formulation. The results, an
electromagnetic gyromagnetic ratio 2 for any spin
together with a non trivial dynamical spin dependent
contact interaction between anyons as well as the spin
dependence of the electromagnetic effective action, agree
with the quantum mechanical formulation.
\vskip 0.25cm
\leftline{DFTUZ 92/15}
\leftline{IPNO/TH 92/99}
\leftline{November 1992}
\vfill\eject

\centerline{\bf 1. Introduction}

Two dimensional theories present some especial features which
can be the basis to understand some phenomena present in planar
systems {\bf [1-3]}. The most remarkable of these features for a quantum
mechanical system is the possibility of excitations with exotic
spin and statistics (anyons) associated to the muticonnectedness
of the rotation group and the configuration space of identical
particles respectively {\bf [4]}.

A dynamical realization of fractional spin-statistics based on
the Aharanov-Bohm effect {\bf [5]} relies on another crucial related
peculiarity of two dimensional theories which is the possibility
to have a gauge invariant first order topological action, the
Chern-Simons (CS) action for a gauge field. At the level of quantum
mechanics it is well understood how the coupling of a CS field to a
bosonic or a fermionic system induces a change in spin and
statistics both at the non-relativistic and relativistic level {\bf [6]}.

In the case of relativistic quantum field theory (RQFT) the
formulation of anyons at the fundamental level is still an open
problem {\bf [7]}. A CS theory as a long distance limit of a topologically
massive U(1) gauge theory with an scalar and/or fermion matter field
has been studied {\bf [8]}. Perturbative calculations as well as different
approaches to the canonical quantization of the theory do not give a
clear picture how anyons are generated in relativistic quantum
field theory {\bf [9]}.

At present the main physical interest in anyon systems is through
its possible connection with some effects in (almost) two dimensional
Condensed Matter Physics systems, in particular the fractional
quantum Hall effect and high temperature superconductivity. From this
point of view it is interesting to extend to a non-relativistic field
theory formulation the dynamical realization of anyon systems in
quantum mechanics. In particular a formulation including an
electromagnetic field is the natural candidate for an effective
theory ${\grave a}$ la Ginzburg-Landau describing the macroscopic effects
associated to the fractional quantum Hall effect and anyon superconductivity
{\bf [10-14]}.

The introduction of the electromagnetic interaction of anyons requires
new ideas compared with the case of bosons or fermions. In a previous work
{\bf [15,16]}
the authors have considered this problem in the context of the CS
formulation of anyons in quantum mechanics. The aim of the present paper is
to extend the formulation to non-relativistic quantum field theory.

The most direct approach corresponds to fix a non-relativistic lagrangian
such that it reproduces all the results of the quantum mechanical
formulation when restricted to the one particle sector. Another approach
takes the gauge invariant renormalizable relativistic quantum field
theory formulation with a bosonic or fermionic field coupled to two
gauge fields (statistical and electromagnetic) as starting point,
and a non-relativistic approximation is considered. Both approaches are
consistent and allow to identify the non-relativistic quantum field theory
formulation of an anyons system in the presence of the electromagnetic
field, which is the main result of this paper.

The article is organized as follows: In section 2 we review the formulation
of a quantum mechanical anyon system based on a spinless (bosonic) and a
spin one half (fermionic) relativistic particle coupled to a CS and
an electromagnetic field; section 3 contains a general discussion of a two
field formulation, which is the natural starting point of a relativistic
quantum field theory formulation, and which allows to understand how the U(1)
gauge theory of a CS formulation of an anyon system is contained in this
general framework. In section 4 we compare the quantum mechanical
 treatment and the associated non-relativistic quantum field theory lagrangian
with previous atempts to formulate a quantum field theory of anyons .
The results of a relativistic quantum field theory with two fields are compared
with the quantum mechanical treatment, and a derivation of the non-relativistic
lagrangian as a non-relativistic limit of a Dirac field coupled to two gauge
fields is presented. In section 5 we give our conclusions and outlook.
\vskip 0.5cm
{\bf 2. $\,\,$Chern-Simons (CS) Formulation of a Quantum Mechanical System in
\indent\indent the
Presence of an Electromagnetic Field }

The simplest construction of fractional spin and statistics is based on
the
introduction of a CS field $ A_\mu$ coupled to a bosonic or fermionic system
through a conserved current $ J_\mu$.
\vfill \eject
The action of the anyon system is

$$I_{an} = I_0 + \int d^3x J_{\mu} (x) A^{\mu} (x) +
{1\over 2\sigma} \int d^3x \,\epsilon_{\mu\nu\rho}A^\mu (x)
\partial^\nu A^\rho (x), \eqno(2.1)$$
where $ I_0 $ is the action of the original bosonic or fermionic system. The
effect of the CS field can be made manifest by integrating over this field
and the resulting effective action is {\bf [17-18]}

$$I_{eff} = I_0 - {\sigma\over 2} \int d^3x\,d^3y \,\,J_\mu (x)
K^{\mu\nu} (x,y) J_\nu (y), \eqno(2.2)$$
where the kernel $ K^{\mu\nu} $

$$K^{\mu\nu} (x, y)= {1\over 4\pi} \epsilon^{\mu\nu\rho}
{{(x - y)}_\rho\over {\vert { x - y} \vert}^3}, \eqno(2.3)$$
is the inverse of the kinetic CS field operator $ \epsilon_{\mu\rho\nu}
\partial^\rho $ acting on a divergenceless vector

$$\int d^3y\,\epsilon_{\mu\rho\nu}{\partial\over \partial x_\rho}
K^{\nu\sigma} (x, y)\,V_\sigma (y)\,=\,V_\mu (x)\,+\,\lq\lq (\partial.V)".
\eqno(2.4) $$

The second term in (2.2) contains the long range interaction responsible of
the change in statistics. The current $ J_\mu$ can include a sum of terms
if one is considering a system of several particles. Then one will have
self interaction diagonal terms and off diagonal particle-particle
interactions.
The second type of terms allow to identify the statistics of the system. In the
relativistic case it is possible to identify the spin induced by the coupling
to the CS field through the contribution of the statistical term to the one
particle action.

If an electromagnetic field is also present then one can also consider a
contribution to $ J_\mu$ from the electromagnetic field, corresponding to a
direct coupling of the statistical and the electromagnetic fields.
\vskip 0.40cm
$\underline {\it 2.a) \,\,Bosonic\,\, anyon.}$

The simplest example of the previous general framework is based
on a free spinless relativistic particle coupled to a CS field, which
corresponds to an action {\bf [15-16]}

$$I_0 = m\int d\tau \sqrt{{\dot X}^2} . \eqno(2.5)$$

For the current $ J_\mu$ one can consider the simplest choice

$$ j_\mu (x) = \int d\tau {\dot X}_\mu \,
\delta^{(3)} ( x -
X(\tau)), \eqno(2.6)$$
which corresponds to a minimal coupling, but nothing prevents to add a magnetic
coupling to the statistical field

$$J_\mu (x) = j_\mu (x)\,+\,\mu_{st}\,\epsilon^{\mu\rho\nu}\partial^\rho j^\nu
(x), \eqno(2.7)$$
where the coefficient $\mu_{st} $ is the statistical magnetic moment. The
effective action for the anyon system will be

$$\eqalignno{I_{eff} =
 &I_0 - {\sigma\over 2} \int d^3x\,d^3y \,\,j_\mu (x)
K^{\mu\nu} (x,y) j_\nu (y) \cr
 &-\,\sigma \mu_{st} \int d^3x\,j_\mu (x) j^\mu (x)\,
-{\sigma\over 2}\mu_{st}^2\,\int d^3x\, \epsilon_{\mu\rho\nu}j^\mu (x)
\partial^\rho j^\nu (x). &(2.8) \cr}$$

The integral in the second term, which is the Gauss linking number of the
trayectories of two particles when an off diagonal contribution is
considered, leads to identify $\sigma $ as the statistical factor of the
system.
The magnetic term in the current (2.7) involves the kinetic operator of the
CS field and then all the effect of the statistical magnetic coupling reduces
to
a local interaction between anyons (the  last two terms in $I_{eff} $ in (2.8)
).
Although irrelevant for the spin-statistics analysis, the non-trivial local
coupling induced by the statistical magnetic moment can have important
consequences on the dynamics of an anyon system, as it has been discussed in
the literature {\bf [19]}.

The CS formulation should be understood as a limit of the most general gauge
invariant action {\bf [20]}, including a standard Maxwell kinetic term, i.e., a
topologically massive gauge theory in the infinite mass limit. The
topological mass M is identified as the relative coefficient of the Maxwell
and CS terms in the gauge field action

$$\eqalignno{I (A) &= \,{1\over 2\sigma} \int d^3x \,\epsilon_{\mu\nu\rho}A^\mu
(x)
\partial^\nu A^\rho (x) \,-{1\over 4M\sigma}\,\int d^3x\,F_{\mu\nu}(A)
F^{\mu\nu}(A)
\cr &={1\over 2\sigma}\,\int d^3x\,A_\mu\,\epsilon^{\mu\rho\sigma}\partial_\rho
[g_{\sigma\nu}\,-{1\over M}\epsilon_{\sigma\eta\nu}\partial^\eta]A^\nu. &(2.9)
\cr}$$

At large distances ($\ell\gg {1 \over M} $) the CS kinetic term dominates and
the CS formulation is recovered with power corrections in the inverse of the
topological mass. Since the contribution to the kinetic operator from the
Maxwell term involves once more the operator associated to the kernel
$ K^{\mu\nu}
$, the same local terms induced by the statistical magnetic coupling are
generated as a finite topological mass effect to order $ {1 \over M} $

$$\eqalignno{I_{eff}^M\,&= I_{eff}\,-\,\,\,{\sigma \over 2M}\,
\int d^3x\,j_\mu (x) j^\mu (x)\,-\,\,\,{\sigma \over 2M}{\mu_{st}}^2\,
\int d^3x\,j^\mu (x) [\partial_\mu \partial_\nu - g_{\mu\nu} \Box ] j^\nu (x)
\cr
&-\,\,\,{\sigma \over M}\mu_{st}\,\int d^3x\, \epsilon_{\mu\rho\nu}j^\mu (x)
\partial^\rho j^\nu (x) \,+ \,{\cal O}({1\over M^2}). &(2.10) \cr}$$

The possibility to cancel the effects induced by a magnetic coupling and a
topological mass, which corresponds to a relation between the statistical
magnetic moment and the topological mass, $ \mu_{st}= -{1\over M}
$, has been considered recently {\bf [21]}.

Once the quantum mechanical formulation of the anyon system has been fixed
the next step is to introduce the electromagnetic interaction. The simplest
 possibility corresponds to consider the Maxwell action for the electromagnetic
field $B_\mu $ together with the standard minimal coupling to the original
spinless particle

$$I\,=\,I_{an}\,+\,e\,\int d^3x\,j_\mu (x)B^\mu (x)\,+\,I_M (B), \eqno(2.11)$$
where

$$I_M (B)\,=\,-{1\over 4}\,\int d^3x\,F_{\mu\nu}(B) F^{\mu\nu}(B),
\eqno(2.12)$$
but one can easily see that something is missing by considering the particular
case of spin $s=\pm{1\over2} $ which corresponds to an statistical factor
$\sigma=\pm2\pi $. In order to reproduce the well known spin dependent
magnetic coupling of a charged spin one half particle one has to add,
together with the minimal coupling to the spinless particle, a direct coupling
 to the CS field, i.e., an additional contribution $J_\mu (B)$ to the
statistical current in (2.7) depending on the electromagnetic field. The
conservation of the current, as required by gauge invariance, forces to use the
operator $ \epsilon_{\mu\rho\nu} \partial^\rho $ in the new contribution to the
statistical current and then all the associated effects  will reduce to a
local coupling. The requirement to reproduce the electromagnetic magnetic
moment $\mu_{em}=\pm{e\over 2m}$ for $s=\pm{1\over2} $ fixes completely
the coupling of both gauge fields

$$\int d^3x\,J^\mu (B)\,A_\mu\,=\,-{1\over 8\pi}{e\over m}\,
\int d^3x\,F_{\mu\nu}(A) F^{\mu\nu}(B). \eqno(2.13)$$

As a consequence of this coupling one has $$\mu_{em}={e\over m}s, \eqno(2.14)$$
i.e. an electromagnetic gyromagnetic ratio $g= 2 $ for any spin and also higher
derivative spin dependent couplings

$$-{e\over m}({\sigma\over 4\pi})(\mu_{st} +{1\over M})\,j_\mu \partial_\nu
F^{\nu\mu}\,-{e\over 2m}{\sigma\over 4\pi}{\mu_{st}\over M}
\epsilon_{\mu\nu\rho}
j^\mu \Box F^{\nu\rho}+\,{\cal O}({1\over M^2}). \eqno(2.15)$$

The only remaining spin dependent effect appears in the effective action for
the electromagnetic field which takes the form

$$I_{eff} (B)\,=\,I_M (B)\,+\,({\sigma \over 4\pi})({e^2\over 4\pi}){1\over 4m}
\,\int d^3x\,\epsilon^{\mu\nu\rho}F_{\mu\nu}\partial^\sigma F_{\sigma\rho}\,+
\,{\cal O}({1\over M}). \eqno(2.16)$$
\vskip 0.40cm
$\underline{{\it 2.b)\,\, Fermionic\,\, anyons.}}$

As a second example of a quantum mechanical formulation of an anyon system,
one can take as starting point the path integral formulation of a spin
one half particle. Anticonmuting variables are used to describe the spin
degrees of freedom and the action is fixed by local supersymmetry {\bf
[22-23]}. When a
minimal supersymmetric coupling to a CS field is included, the effective
action after integration over the anticonmuting variables gives two
contributions, corresponding to the spin plus and minus one half components,
which can be reduced to the general form (2.1) with

$$J_\mu^\pm\,=j_\mu\,\pm{1\over 2m}\epsilon_{\mu\rho\nu}\partial^\rho j^\nu
.\eqno(2.17)$$

The original spin degrees of freedom induce a magnetic statistical moment
$\mu_{st}=\pm{1\over 2m} $. The local supersymmetry of the action does not
allow
in the present case to introduce a magnetic statistical coupling at the
beginning unless one goes beyond the standard lagrangian formulation including
higher derivatives of the statistical field. The value of $\mu_{st} $ is
directly connected to the gyromanetic ratio of a spin one half particle, which
in the present formulation is a direct consequence of the local supersymmetry
of
the action.

The general discussion of the bosonic anyon formulation can be translated
directly to the fermionic case with the only difference that the statistical
magnetic moment, which in the previous case was arbitrary, is now fixed,
$\mu_{st}=\pm{1\over 2m} $. On the other hand the spin induced by the
coupling to the CS field, identified from the coefficient of the diagonal
statistical interaction $({\sigma\over 4\pi}) $, has to be added now to the
original spin $\pm{1\over2} $.

The first step in the introduction of the electromagnetic field in this case is
to couple it to the spinning particle in a supersymmetric way which, after
integration over the fermionic degrees of freedom, leads to

$$I^\pm\,=\,I_{an}^\pm\,+\,e\,\int d^3x\,J_\mu^\pm B^\mu\,+\,I_M(B),
\eqno(2.18)$$
where $I_{an}^{\pm}$ is a particular case of the general action of the
CS anyon system with the current $J_\mu^{\pm} $ as the statistical current.

Once more one can see that this is not the final answer by considering the
case $\sigma=-2\pi $ ( $\sigma=2\pi $) in the action $I^+ $ ($I^- $) which
corresponds to a spinless system obtained from the fermion. A magnetic coupling
to a scalar field is forbiden by renormalizability.If one wants to
get $\mu_{em}= 0 $ at the level of quantum mechanics an additional
contribution to the electromagnetic magnetic moment is required and the only
remaining possibility is to add a contribution to the statistical current
involving the electromagnetic field, i.e., a direct coupling of both gauge
fields. In the fermionic formulation the CS field can be thought together with
the anticonmuting variables as a way to describe the spin degrees of freedom of
the system. Since the local supersymmetry of the spinning
action leads to consider a coupling of the electromagnetic field to the
anticonmuting variables, it is natural to have at the same time a coupling
of the electromagnetic field to the CS field. This coupling can be determined
from the electromagnetic moment analysis and it coincides with the coupling
required in the bosonic formulation, finding once more a gyromagnetic
ratio $g = 2 $ for the (fermionic) anyon for any spin.
\vskip 0.50cm
{\bf 3. The Electromagnetic Interaction of Anyons from a General Point
 $\,\,\,\,\,\,\,\,\,\,\,\,$ of View.}

The problem of the electromagnetic interaction of anyons involves two gauge
fields: the statistical field $A_\mu$ and the electromagnetic field $B_\mu$. If
at the
begining we see both of them at the same level (the coupling required in the
quantum mechanical discussion points in this direction) we will understand the
problem in a more general way. From this point of view we have to consider the
construction of a U(1)$\times$U(1) gauge theory coupled to a matter field
trough
a conserved current. Then we will start with a lagrangian
$${\cal L}\,=\,{\cal L}_{gauge}(A_1,A_2)\,+\,q_i J_\mu A_i^\mu\,+\,{\cal L}
_{matter}, \eqno(3.1)$$
where $A_i^\mu $ (i=1,2) are the two gauge fields coupled to the matter field
through the conserved current $J^\mu$ with charges $q_i$ and ${\cal L}_{gauge}$
is the
most general lagrangian for the fields $A_i$ in 2+1 dimensions. It will involve
two independent quadratic forms
$${\cal L}_{gauge}(A_1,A_2)\,=\,\alpha_{ij}\,{\cal L}_{CS}(A_i,A_j)\,+
\,\beta_{ij}\,{\cal L}_{M}(A_i,A_j), \eqno(3.2)$$
where
$$\eqalignno{&{\cal L}_{CS}(A_i,A_j)\,={1\over 2}A_i^\mu \epsilon_{\mu\nu\rho}
\partial^\nu A_j^\rho, \cr &{\cal L}_{M}(A_i,A_j)\,=\,-{1\over 4}
F_{\mu\nu}(A_i) F^{\mu\nu}(A_j). &(3.3)\cr}$$

Of course we can perform a linear transformation in the fields $A_i$ to
diagonalize simultaneously the two quadratic forms. So we can consider
without lost of generality a gauge lagrangian

$${\cal L}_{gauge}(A_1,A_2)\,=\,\alpha_i \,{\cal L}_{CS}(A_i,A_i)\,+
\,\beta_i \,{\cal L}_{M}(A_i,A_i). \eqno(3.4)$$

What can we say about $\alpha_i$ and $\beta_i$ ?. First $I_{gauge}=
\int d^3 x {\cal L}_{gauge}$ must be bounded from below and this implies for
the
Maxwell quadratic form to be positive definite, that is, $\beta_i>0$.
On the other hand we want only one statistical field $A$; otherwise one
would not recover the standard electromagnetic interaction. Therefore only
one linear combination of the two fields $A_i$ can appear in the CS like
terms, i.e., $\alpha_2=0$ and $A=A_1$ up to normalization factors. In this
way we have identified the statistical field without ambiguity. What about
the electromagnetic field?. All we can say is that it must be a linear
combination of the fields $A_i$ independent of $A$, that is $B=A_2+\gamma A_1$
again up to normalization factors. The parameter $\gamma$ reflects the
ambiguity in determining the electromagnetic field and it fixes how the
electromagnetic U(1) interaction is embeded into the original U(1)$\times$U(1)
symmetry. In terms of the fields $A_\mu$ and $B_\mu$ the gauge lagrangian will
be

$${\cal L}_{gauge}\,=\,a{\cal L}_{CS}(A,A)\,+\,b_{st}{\cal L}_{M}(A,A)
\,+\,b_{em}{\cal L}_{M}(B,B)\,+\,c{\cal L}_{M}(A,B), \eqno(3.5)$$

with

$$a=\alpha_1\,,\,\,\,\,b_{st}=\beta_1 +\gamma^2 \beta_2\,,\,\,\,\,
b_{em}=\beta_2\,,\,\,\,\,c=-2\gamma \beta_2. \eqno(3.6)$$

Therefore, although we can diagonalize the gauge action in general we will
have to consider the possibiblity of an interaction between the statistical
and electromagnetic fields through a Maxwell like term. From this point of
view the result of the previous section was a determination of this interaction
,i.e., an identification of the way the electromagnetic interaction is inside
the total U(1)$\times$U(1) gauge interaction through the analysis of the
magnetic interaction of the anyon
system. If one uses the remaining rescaling freedom of the fields to get
$b_{em}=1$ and
$q_{st}=1$ then the gauge field lagrangian in (3.5) reproduces the gauge field
lagrangian identified in section 2 when

$$a={1\over \sigma}\,,\,\,\,\,b_{st}={1\over M\sigma}\,,\,\,\,\,
c={1\over 2\pi}{e\over m}. \eqno(3.7)$$

Let us finish this section with some remarks about ${\cal L}_{gauge}$ and the
distinction between A and B. We know that in relativistic quantum field
theory there will be matter fluctuations associated to the creation and
anihilation of particle-antiparticle pairs which generate radiative
corrections to the gauge lagrangian. Therefore, at the level of relativistic
quantum field theory we have to think of ${\cal L}_{gauge}$ as the bare
lagrangian
while in the other cases (quantum mechanics and non-relativistic quantum
field theory) ${\cal L}_{gauge}$ will include the effect of the radiate
corrections
which are not incorporated in these approximations. Of course one has to
consider only the corrections to the terms already present in the original
lagrangian (that is, quadratic in the fields and with no more than two
derivatives).
This means for example that (in both cases, bosonic and fermionic) there
will be corrections for the kinetic Maxwell term for both fields $B_\mu$
and $A_\mu$. The
correction for the electromagnetic field will just translate into a
redefinition
of the electric charge. However, in the case of the statistical field it
implies
a positive redefinition of the topological mass $M$ of the order of the mass
of the particle $m$

$${1\over M}={1\over M_0}\,+\,{1\over 24\pi}{1\over m}, $$
$${1\over M}={1\over M_0}\,+\,{1\over 12\pi}{1\over m}, \eqno (3.8)$$
for the bosonic and fermionic case respectively, where $M_0 $ is the bare
topological mass. This effect prevent us to take the ideal limit $M\to\infty$
while keeping the particle mass finite. This means that one can only have an
anyon at large distances compared with ${1\over m}$, where the CS term for A
will dominate over the Maxwell term. Note that this limit is consistent with
the validity of the quantum mechanical and nonrelativistic field theory
approximations.

At the same level there will be corrections to the Maxwell interaction
between $A_\mu$ and $B_\mu$ and therefore the coupling determined at the level
of
quantum mechanics includes the effect of radiative corrections. This means
that the coupling at the level of the relativistic quantum field theory
lagrangian will not be the same.

What about the CS like terms? The same discussion can be applied to this case.
In the bosonic case, due to the parity invariance of the free theory, there
will be no correction induced at this level from radiative corrections and
${\cal L}_{gauge}$ will coincide at all levels. In the fermionic case, the
redefinition
of the CS like terms induced by radiative corrections is a regularization
dependent effect.
\vskip 0.25cm
\centerline{\bf 4. Non-Relativistic Field Theory of Anyons.}
\vskip 0.25cm

$\underline{{\it 4.a) \,\,From\,\, quantum\,\, mechanics\,\, to\,\,
non-relativistic\,\, field\,\, theory.}}$

The most direct approach to the non-relativistic field theory formulation of an
anyon system is based on the identification of a non-relativistic lagrangian
which, when restricted to the one particle sector, reproduces the quantum
mechanical formulation of the electromagnetic interaction of anyons. The most
natural formulation, the fermionic one, leads to consider a particle coupled
to two fields, a CS field and an electromagnetic field, with a spin dependent
magnetic coupling corresponding to $\mu_{st}={1\over 2m}$ and
$\mu_{em}={e\over m}\, s$. At the level of quantum field theory the fermionic
anyon formulation translates into a lagrangian with an anticonmuting field
$\Phi$ coupled to a CS field

$${\cal L}_{an}\,=\,\Phi^+ iD_0 \Phi \,-\,{1\over 2m}\Phi^+
{\vec D}^2 \Phi\,+\,{i\over 2m}\Phi^+ [D_1 ,D_2] \Phi \,+\,
{1\over 2\sigma} \epsilon_{\mu\nu\rho}A^\mu
\partial^\nu A^\rho, \eqno(4.1)$$
where $D_\mu=\partial_\mu-iA_\mu$ is the covariant derivative involving the
statistical field. The sum of the first two terms in (4.1) is the standard non
relativistic lagrangian whose Euler-Lagrange equation is the Schroedinger
equation minimally coupled to a field $A_\mu$ and the third term is necessary
in order to reproduce the statistical magnetic moment $\mu_{st}={1\over 2m}$
when considering the one particle sector. The presence of a non-minimal
magnetic coupling is in agreement with previous work.

Recently the non-relativistic (bosonic) field theory for the second quantized
N-body system of point particles with Chern-Simons interactions has been
constructed {\bf [24]}
. It includes a quartic interaction which alternatively can be understood as
a magnetic field-charge density interaction whose intensity can be fixed in
order to render the system self-dual allowing to identify regular, static
classical soliton solutions {\bf [25]}. The lagrangian (4.1) seems to be the
fermionic version of this construction and the magnetic term introduced to
render the system self-dual is nothing but the field theoretic translation of
the
statistical magnetic moment $\mu_{st}={1\over 2m}$ identified in the quantum
mechanical treatment \footnote {$^\dagger$}{Note that the specific value of the
magnetic interaction which renders the system self-dual corresponds to a
supersymmetric hamiltonian {\bf [26]}. We thank A. Comtet and R. Iengo by
drawing our attention to this reference.}.

In a related work the possibility to include a self-interaction in the second
quantized bosonic hamiltonian of anyons which takes care of operator ordering
and operator product singularities has been considered {\bf [27]}. When the
self
interaction is fixed in order to reproduce the spectrum of the two anyon
system in the background of a uniform magnetic field one finds that a repulsive
local interaction is required with the same intensity as the local atractive
interaction induced by the statistical magnetic moment $\mu_{st}={1\over 2m}$
in the fermionic formulation.

Once the non-relativistic field theoretic version of the quantum mechanical
fermionic anyon system has been identified one can try to include
the coupling to the electromagnetic field at this level. The first step
corresponds to add the Maxwell lagrangian for the electromagnetic field and
to include the electromagnetic field in the covariant derivative which appeared
in ${\cal L}_{an} $, $D_\mu \to {\cal D}_\mu=\partial_\mu-iA_\mu-ieB_\mu$ .
This reproduces,
in the one particle sector, the result of a minimal supersymmetric coupling
in the spinning particle leading to $\mu_{st}={1\over 2m}$ ,
$\mu_{em}={e\over 2m}$ for any
spin. Once more if one wants to reproduce the absence of an electromagnetic
magnetic coupling in the spinless case a modification of the lagrangian is
required. Since the second quantized formulation reproduces the quantum
mechanical discussion in the one particle sector we can use directly the
results in section 2. The conclusion is that one should add a direct coupling
between the statistical and the electromagnetic fields as given by (2.9) and
the non-relativistic fermionic field theory lagrangian for the anyon system
in the presence of the electromagnetic field is

$$ \eqalignno{{\cal L}\,=\,&\Phi^+ i{\cal D}_0 \Phi \,-\,{1\over 2m}\Phi^+
{\vec {\cal D}}^2 \Phi\,+\,{i\over 2m} \Phi^+ [{\cal D}_1 ,{\cal D}_2] \Phi
\,+\,
\cr & +{1\over 2\sigma} \epsilon_{\mu\nu\rho}A^\mu  \partial^\nu A^\rho \,-\,
{1\over 4} F_{\mu\nu}(B) F^{\mu\nu}(B) \,-\,
{1\over 8\pi}{e\over m} F_{\mu\nu}(A) F^{\mu\nu}(B), &(4.2)\cr}$$
with ${\cal D}_\mu=\partial_\mu-iA_\mu-ieB_\mu$.

The non-relativistic field theoretic version of the quantum mechanical anyon
system can be obtained along the same lines and the only difference is that
the matter field $\Phi$ satisfies canonical conmutation relations in this case
and that the magnetic term involving the statistical field
$i\Phi^+[D_1,D_2]\Phi$
would have an arbitrary coefficient corresponding to the arbitrariness in
$\mu_{st}$ in the bosonic quantum mechanical discussion.

$\underline{{\it 4.b)\,\, Non-relativistic\,\, limit\,\, of\,\,
relativistic\,\,
 quantum\,\, field \,\,theory.}}$

Another approach to the non-relativistic field theory of anyons is based on
the identification of the field theoretical version of the relativistic anyon
system followed by a non-relativistic approximation. A relativistic quantum
field theory formulation of anyons presents additional problems, associated
to the non localized character of the statistical current, which make the
identification of operators associated to the fractional spin-statistics
excitations still an open question {\bf [28]}. This is the reason why a
discussion
of anyons in a relativistic quantum field theoretic formulation presents
ambiguities at a fundamental level. The basic formulation is based on a scalar
or a fermionic field minimally coupled to a CS field. Recently it has been
shown {\bf [29]} that a one loop perturbative analysis of the statistical field
fluctuations generates a magnetic coupling to both the statistical and
electromagnetic fields. This coupling leads, at the level of Green functions of
the elementary
fields, to a gyromagnetic ratio $g=2$; it is not clear for us how
this result is related with the gyromagnetic ratio of an anyon. We think this
is the explanation why the statistical magnetic moment induced by radiative
corrections $\mu_{st}={1\over m}s$ differs from the anyon statistical
magnetic moment identified in the quantum mechanical discussion
$\mu_{st}={1\over 2m}$ for any spin. The same criticism applies to the
analysis of short range interaction as deduced from the two-particle
scattering amplitude in relativistic quantum field theory and its relation
to the short range interaction between anyons induced by the statistical
magnetic moment.

One can formally reproduce the fermionic anyon discussion of quantum mechanics
starting from the Dirac action with a statistical gauge field and defining a
current $J_\mu$ by

$${\displaystyle{\bar \Psi}}\gamma_\mu \Psi \,=\,J_\mu \,+\,{1\over
2m}\epsilon_{\mu\rho\nu}
\partial^\rho J^\nu, \eqno(4.3)$$

Note that the effective statistical current $J_\mu $ is a non-local expression
in terms of the fundamental fields as expected for the anyon current
$$J_\mu (x)\,=\, \int \,d^3y\,G_{\mu\nu}(x,y)\,{\bar \Psi}\gamma^\nu
\Psi (y) \,,\eqno(4.4)$$
with
$$( g_{\mu\nu}\,+\,{1\over 2m} \epsilon_{\mu\rho\nu}{\partial \over \partial
x_\rho}) G^{\nu\sigma}(x,y)\,=\,\delta_\mu ^\sigma \delta^3(x-y) \,,
\eqno(4.5)$$
but we do not know what are the consequences of this formal connection of the
quantum mechanical and relativistic quantum field theory formulations.

One can go further in this connection by going to the non-relativistic limit
and considering a sector with a given number of particles. In order to do that
one first factorizes the dominant time dependence of the Dirac field $\Psi$

$$\Psi \,=\,e^{-imt}\,\Phi \,=\,e^{-imt}\,\pmatrix {\chi \cr \eta \cr},
\eqno(4.6)$$
and using the 2+1 dimensional representation of the Clifford algebra
$$\gamma^0 =\sigma^3 =\pmatrix{ 1&0\cr 0&-1\cr }\,\,\,\,\,\,\,\,,
\gamma^1 =i\sigma^1 =\pmatrix{ 0&i\cr i&0\cr }\,\,\,\,\,\,\,\,,
\gamma^2 =i\sigma^2 =\pmatrix{ 0&1\cr -1&0\cr }\,\,\,\,\,\,\,\,,\eqno(4.7)$$
the Dirac action coupled to the statistical field reads
$$ \eqalignno{{\cal L}_D \,&=\,{\bar \Psi}i\gamma^\mu (\partial_\mu -
iA_\mu) \Psi \,-m {\bar \Psi} \Psi \,= \cr &=\,{\bar \chi}iD_0 \chi \,+\,
{\bar \eta}\,(-iD_0 -2m)\, \eta \,+ {\bar \chi}\,(-D_1 + iD_2)\, \eta \,+ \,
{\bar \eta}\,(-D_1 - iD_2)\, \chi, & (4.8) \cr}$$

{}From this expression one can identify the combinations

$$\eta^\prime \,=\,\eta \,+\,(iD_0+2m)^{-1}\,(D_1 +iD_2)\,\chi $$
$${\bar \eta}^\prime \,=\,{\bar \eta} \,+\,{\bar \chi}\,(D_1 -iD_2)\,(iD_0+2m)
^{-1}\, \eqno(4.9)$$
which diagonalize the Dirac action

$${\cal L}_D \,=\,{\bar \chi}iD_0 \chi \,+\,{\bar \chi}\,(D_1 -iD_2)\,(iD_0+2m)
^{-1}\,(D_1 +iD_2)\,\chi\,+{\bar \eta}^\prime \,(-iD_0 -2m)\, \eta^\prime \,.
\eqno(4.10)$$

In the non-relativistic limit one can consider a smoth time dependence for
$\chi,{\eta^\prime}$ and assuming ${\vert A_0 \vert} \ll 2m$ one has

$${\cal L}_{N.R.} \,\approx\,{\bar \chi}iD_0 \chi \,+\,{1\over 2m}{\bar
\chi}\,(D_1 -iD_2)
\,(D_1 +iD_2)\,\chi\,-\,2m\,{\bar \eta}^\prime \,\eta^\prime \,+\,
{\cal L}_{eff}(A), \eqno(4.11)$$
where ${\cal L}_{eff}(A)$ is the original action of the statistical field plus
the effect of the quantum fluctuations of the matter field which include a
renormalization of the statistical factor.

At this level the variable $\eta^\prime$ trivially factorizes and one gets the
non-relativistic lagrangian for the anticonmuting field $\chi$ which using
the identity

$$(D_1 -iD_2)\,(D_1 +iD_2)\,={\vec D}^2 \,+\,i[D_1,D_2], \eqno(4.12)$$
can be identified as the sum of the naive non-relativistic lagrangian
minimally coupled to the statistical field plus a magnetic term with
$\mu_{st}={1\over 2m}$. Then the non-relativistic limit of the relativistic
quantum field theory action reproduces the result previously derived from
the quantum mechanical fermionic formulation.

In the presence of the electromagnetic field the relativistic lagrangian
will be

$${\cal L}_R \,=\,{\bar \Psi}\,(i \dsl + \Asl +e \Bsl -m) \,\Psi \,+\,
{\cal L}(A,B), \eqno(4.13)$$
where ${\cal L}(A,B)$ is the more general renormalizable gauge invariant action
for two fields as discussed in section 3. One can repeat the derivation
of the non-relativistic limit by including the electromagnetic field in the
covariant derivative. Let us assume a lagrangian ${\cal L}(A,B)$ for the two
fields such that when the matter field quantum fluctuations are included
one obtains

$$\eqalignno{{\cal L}_{eff}(A,B)\,&\approx\,{1\over 2\sigma}
\,\epsilon_{\mu\nu\rho}
A^\mu \partial^\nu A^\rho \,-{1\over 8\pi}{e\over m}
F_{\mu\nu}(A) F^{\mu\nu}(B)\,-\,{1\over 4} F_{\mu\nu}(B) F^{\mu\nu}(B) \,\cr
&-\,
{1\over 4} {1\over \sigma M} \, F_{\mu\nu}(A) F^{\mu\nu}(A) \,+\,
higher\,\,\,derivative\,\,\,terms\,, & (4.14) \cr}$$
which is the two field lagrangian required at the quantum mechanical level.
One can repeat the derivation of the non-relativistic lagrangian including
the electromagnetic field and the result in {\bf [15-16]} is obtained as a
direct
consequence of the gauge invariant coupling of the fields A and B in the
relativistic lagrangian. The only part which is not understood at this level
is what is the principle (symmetry) which fixes the required two gauge field
lagrangian.

The magnetic term induced by the coupling of the CS and electromagnetic fields
in ${\cal L}_{eff}$ can be easily identified if one introduces the combination

$${\tilde A}_\mu \,=\,A_\mu \,-\,{e\over m}{\sigma\over 4\pi}
\epsilon_{\mu\nu\rho} \partial^\nu
B^\rho, \eqno(4.15)$$
which diagonalizes ${\cal L}_{eff}$ (up to higher derivative terms). This
redefinition induces a magnetic coupling of the electromagnetic field and the
matter current $J_\mu $

$$\int d^3x J_{\mu}A^\mu \,=\,\int d^3x J_{\mu} {\tilde A}^\mu \,+\,{e\over m}
{\sigma\over 4\pi}
\,\int d^3x J_{\mu}\,\epsilon^{\mu\nu\rho} \partial_\nu B_\rho,
\eqno(4.16)$$

This argument can be applied whenever the coupling to the statistical field
is linear in the CS field, as in the quantum mechanical discussion, and it
can also be applied to the relativistic quantum field theory lagrangian where
after using the identity

$$\epsilon_{\mu\nu\rho}\,\gamma^\mu \,=\,{i\over 2}[\gamma_\nu ,\gamma_\rho
]\,=\,
\sigma_{\nu\rho} \,, \eqno(4.17)$$

one has

$${\cal L}_R \,=\,{\bar \Psi}\,(i \dsl + {\tilde {\Asl}} +e \Bsl -m) \,\Psi
\,+\,{e\over m}{\sigma\over 4\pi}{\bar \Psi}\sigma_{\mu\nu} \Psi \,
F^{\mu\nu}(B) \,+\,
{\cal L}(A,B). \eqno(4.18)$$

We can identify a Pauli magnetic interaction as a consequence of the
redefinition
of the statistical field giving an anomalous magnetic moment contribution which
in the non relativistic limit leads to $\mu_{em}={e\over m}({1\over 2} +
{\sigma\over 4\pi})$.

An attempt to extend the bosonic anyon formulation starting with a
relativistic
quantum field theory is based on a complex scalar field coupled to a CS field.
The gauge invariant renormalizable lagrangian contains new parameters
associated to the self interaction of the scalar field with no analog in
the fermionic formulation. In the non-relativisitic limit the self coupling
translates into a local quartic term which using the equations of motion of
the CS field can be alternatively represented as a magnetic coupling. The
arbitrariness of the quantum mechanical bosonic formulation in $\mu_{st}$
reappears once more in field theory through the self coupling of the scalar
field.
\vskip 0.50cm
\centerline{\bf 5. Conclusions and Outlook}

A detailed discussion of a relativistic anyon system at the quantum
mechanical level has been presented. The non-trivial dynamical effects
associated to a statistical magnetic moment have been identified in a
bosonic and in a fermionic formulation. The resulting local interactions,
which differ from previous analysis in relativistic quantum field theory,
can be very important for the study of a pairing anyon mechanism and its
possible application to superconductivity.

The introduction of the electromagnetic field in the CS construction has
been studied including the determination of a direct coupling between
both fields in order to obtain a consistent spin dependence of the
electromagnetic interaction at the quantum mechanical level. Together
with a non trivial spin dependence of the effective action of the
electromagnetic field, the main result is that an anyon has a gyromagnetic
ratio g=2 for any spin. This result seems to be a non-trivial generalization
to a two dimensional system of a result obtained recently {\bf [30]} in three
spatial
dimensions, where it has been found that at tree level g=2 is the natural
value for the gyromagnetic ratio for any spin (integer and half integer) in
order to have a smooth massless limit which is behind the unitarity of the
tree level scattering amplitude. This suggests a possible derivation of the
anyon magnetic moment from first principles.

The non-relativistic lagrangian corresponding to the second quantized
fermionic formulation which reproduces in the one particle sector the
quantum mechanical treatment of a fermionic anyon system has been identified.
It includes a non-minimal coupling to the statistical and electromagnetic
fields and a direct coupling between both fields as new ingredients with
respect to previous formulations. The identification of these new terms is
the main result of this paper. An explicit derivation of the non minimal
coupling as a result of the non-relativistic reduction of the Dirac lagrangian
with a minimal coupling to both fields has been presented.

The implications of the new terms in the non-relativistic quantum
field theoretical lagrangian of the anyon system presented in this paper,
when considered as the lagrangian of the effective theory
of the fractional quantum Hall effect as well as a starting point for the
study of anyon superconductivity, deserve further investigation.

{\bf Acknowledgments}. We would like to thank A.Comtet, R.Iengo, H.B.Nielsen
and M. Valle for useful discussions. The work of J.G. has been partially
supported by CNRS.

\vskip 0.5cm
\centerline{\bf References}
\vskip 0.25cm
\item{\bf [1]} See e.g. E. Fradkin, {\it Field Theories of Condensed Matter
Systems}, Addison-Wesley 1991.
\item{\bf [2]} J.W. Lynn, {\it High Temperature Superconductivity},
Springer-Verlag Inc., New York, 1990.
\item{\bf [3]} R. Prange and S. Girvin, {\it The Quantum Hall Effect},
Springer-Verlag Inc., New York, 1987.
\item{\bf [4]} The original references are collected in  F. Wilczek, {\it
Fractional Statistics and Anyon
Superconductivity}, World Scientific, 1990.
\item{\bf [5]} Y. Aharonov and D. Bohm, Phys. Rev. {\bf 115}(1959)485.
\item{\bf [6]} Recents reviews on this points and other extensions are
S. Forte, Rev. Mod. Phys. {\bf 64}(1992)193.; K. Lechner
 and R. Iengo, Phys. Rep {\bf 213}(1992)179.
\item{\bf [7]} F. Wilczek and A. Zee, Phys. Rev. Lett. {\bf 51}(1983)2250;
 G.W.
Semenoff,  Phys. Rev. Lett. {\bf 61}(1988)517;
G.W. Semenoff and P.Sodano, Nucl.Phys. {\bf B328}(1989)753;
S. Forte and T. Jolicoeur,
Nucl. Phys. {\bf B350}(1991)589.
\item{\bf [8]} S. Deser, R. Jackiw and S. Templeton, Ann. of Phys. (N.Y.)
{\bf 140}(1982)372
\item{\bf [9]} D. Boyanovsky, Phys. Rev. {\bf D42}(1990)1179; D. Boyanovsky,
E.T. Newman and C. Rovelli, Phys. Rev. {\bf D45}(1992)1210; R. Banerjee,
Phys. Rev. Lett. {\bf 69}(1992)17.
\item{\bf [10]} S.M. Girvin and A.H. MacDonald , Phys. Rev. Lett. {\bf
 58}(1987)1252.
\item{\bf [11]} E. Rezayi and F.D.M. Haldane, Phys. Rev. Lett.
{\bf 61}(1988)1985.
\item{\bf [12]} S. Zhang, T.H. Hanson and S. Kivelson, Phys. Rev. Lett. {\bf
62}
(1989)82.
\item{\bf [13]} N. Read, Phys. Rev. Lett. {\bf 62}(1989)86.
\item{\bf [14]} J. Fr\"olich and T. Kerler, Nucl. Phys. {\bf B354}(1991)369.
\item{\bf [15]} J.L. Cort\'es, J. Gamboa and L. Vel\'azquez, Phys. Lett. {\bf
B286}
(1992)105.
\item{\bf [16]}J.L. Cort\'es, J. Gamboa and L. Vel\'azquez, {\it A $U(1)$ Gauge
Theory
for Anyons}, DFTUZ 91/30 preprint. Nucl. Phys. B (in press)
\item{\bf [17]} A.M. Polyakov, Mod. Phys. Lett. {\bf A3}(1988)325.
\item{\bf [18]} S. Forte, Int. J. Mod. Phys. Lett. {\bf A7}(1992)1025.
\item{\bf [19]} I.I. Kogan, Phys. Lett. {\bf B262}(1991)83.
\item{\bf [20]} K. Shizuya and H. Tamura, Phys. Lett. {\bf B252}(1990)412.
\item{\bf [21]} J. Stern, Phys. Lett. {\bf B265}(1991)119.
\item{\bf [22]} L. Brink, P. Howe, P. di Vecchia, S. Deser and B. Zumino,
Phys. Lett. {\bf B64}(1976) 435.
\item{\bf [23]} F.A. Berezin and M.S. Marinov, Ann. of Phys. (N.Y.){\bf 104}
(1977)336.
\item{\bf [24]} R. Jackiw and S.Y. Pi, Phys. Rev. {\bf D42}(1990)3500.
\item{\bf [25]} R. Jackiw and S.Y. Pi, Phys. Rev. Lett. {\bf 64}(1990)2969.
\item{\bf [26]} S.M. Girvin, A.H. Mac Donald, M.P.A. Fisher, S.J. Rey and J.P.
Sethna, Phys. Rev. Lett. {\bf 65}(1990)1671.
\item{\bf [27]} Z.F. Ezawa, M. Hotta and A. Iwasaki, Nisho-IR (Tohoku) 1991,
preprint.
\item{\bf [28]} J. Fr\"olich and P.-A. Marchetti, Lett. Math. Phys. {\bf
16}(1988)347.
\item{\bf [29]} I.I. Kogan and G.W. Semenoff, Nucl. Phys. {\bf B368}(1992)718.
\item{\bf [30]} S. Ferrara, M. Porrati and V.L. Teledgi, Phys. Rev. {\bf D46}
(1992)3529.
\end